\documentclass[prl,aps,twocolumn,groupedaddress,floats,showpacs,final,superscriptaddress]{revtex4-2}
\usepackage[latin3]{inputenc}
\usepackage[makeroom]{cancel}
\usepackage{graphicx}
\usepackage{amsmath}
\usepackage{amsfonts}
\usepackage{amssymb}
\usepackage{chngcntr}
\usepackage{color}
\usepackage[left=2cm,right=2cm,top=2cm,bottom=2cm]{geometry}

\usepackage{graphicx}
\usepackage{dcolumn}
\usepackage{bm}
\usepackage{simplewick}
\usepackage{array}
\usepackage{appendix}

\RequirePackage[
   hyperindex,colorlinks,bookmarksnumbered,
   plainpages=true,pdfstartview=FitH]{hyperref}
\hypersetup{linkcolor=blue,urlcolor=blue,citecolor=blue}
\usepackage{hyperref}

\definecolor{purple}{rgb}{0.5,0,0.6}

\begin{document}

\title{Photoinduced Anomalous Supercurrent Hall Effect}



\author{A.~V.~Parafilo}
\affiliation{Center for Theoretical Physics of Complex Systems, Institute for Basic Science (IBS), Daejeon 34126, Korea}


\author{V.~M.~Kovalev}
\affiliation{Rzhanov Institute of Semiconductor Physics, Siberian Branch\\ of Russian Academy of Science, Novosibirsk 630090, Russia}
\affiliation{Novosibirsk State Technical University, Novosibirsk 630073, Russia}

\author{I.~G.~Savenko}
\affiliation{Department of Physics, Guangdong Technion -- Israel Institute of Technology, 241 Daxue Road, Shantou, Guangdong, China, 515063}
\affiliation{Technion -- Israel Institute of Technology, 32000 Haifa, Israel}
\affiliation{Guangdong Provincial Key Laboratory of Materials and Technologies for Energy Conversion, Guangdong Technion--Israel Institute of Technology, Guangdong 515063, China}

\date{\today}


\begin{abstract}
We predict a photoinduced Hall effect in an isotropic conventional two-dimensional superconductor with a built-in supercurrent exposed to a circularly-polarized light. 
This second-order with respect to the electromagnetic field amplitude effect occurs when the frequency of the field exceeds the double value of the superconducting gap. It reveals itself in the emergence of a Cooper-pair condensate flow in the direction transverse to the initial built-in supercurrent, which arises to compensate for the light-induced electric current of quasiparticles photoexcited across the gap. 
The initial supercurrent breaks both the time-reversal and inversion symmetries, while the presence of dilute disorder in the sample provides the breaking of the Galilean invariance. 
We develop a microscopic theory of the supercurrent Hall effect in the case of weak disorder and show, that the Hall supercurrent is directly proportional to the quasiparticle recombination time, which can acquire large values.
\end{abstract}

\maketitle

\textit{Introduction.---} 
The measurement of optical response in superconductors is a powerful experimental technique to explore their quantum properties, which are in the focus of  research for more than fifty years~\cite{RevModPhys.46.587, 10.1007/978-1-4684-1863-7_9, RevModPhys.77.721,Arseev_2006, Charnukha_2014,dressel, PhysRevB.98.064502, PhysRevLett.123.217004, PhysRevLett.124.207002}. 
Despite such an extensive period of time, the study of interaction of electromagnetic (EM) fields with superconductors remains a challenging topic since, as it is known, superconducting (SC) samples usually expel external EM fields~\cite{Tinkham}. 
Beside fundamental importance, the research on light-controlled transport of Cooper pairs is aimed at applications~\cite{JiangHuNature2022, PhysRevLett.124.087701}.
Examples of possible (but not yet implemented) light-matter interaction phenomena in superconductors include various nonlinear and higher-order response effects \cite{eliash1,eliash2,PhysRevB.99.224511}, in particular, electric field-induced enhancement of SC properties \cite{Eliashb,PhysRevB.97.184516}, giant second-harmonic generation under supercurrent injection~\cite{PhysRevLett.125.097004}, and light-mediated superconductivity~\cite{PhysRevLett.104.106402, Sun_20212DMater, SunNJP2021}, to name a few. 
Another route, which we inspect in this Letter, would be a photoinduced anomalous Hall effect -- a possibility to manipulate a dc supercurrent flow by utilizing the Hall-like response.

Anomalous Hall effect in non-SC samples represents a stationary transport phenomenon, which constitutes the emergence of a transverse component of electric current in the absence of an external magnetic field~\cite{RevModPhys.82.1539}.
The examples are the spin Hall effect, where spin-orbit interaction plays the role of the magnetic field, the valley Hall effect~\cite{XiaoVHE, Mak1489, PhysRevLett.122.256801, Jin893} in two-dimensional (2D) Dirac materials~\cite{RefTMDs1, RefTMDs2, RefLiu2019, Kovalev2018NJP}, and
the photoinduced anomalous Hall effect, actively studied in various systems~\cite{RefMclver2020}.
Is it possible to find a setup for such an effect to involve Cooper pairs?

Before answering this question, let us briefly review 
the optical response theory in superconductors.
As it is known, in clean single-band  Bardeen-Cooper-Schrieffer (BCS) superconductors~\cite{PhysRev.108.1175}, the presence of particle-hole and inversion symmetries does not allow for momentum-conserving optical transitions~\cite{PhysRev.108.1175,Mahan,Tinkham}.
The optical excitations under inversion symmetry can occur if account for either impurity scattering \cite{PhysRev.111.412,Mahan} or multiband structure~\cite{NagaosaOptRe2021,PhysRevB.106.214526}. 
The first theoretical analysis of the dynamical conductivity of superconductors exposed to EM fields with the frequency exceeding the SC gap belongs to Mattis and Bardeen~\cite{PhysRev.111.412}, who considered the `dirty case'. 
They have shown, that in the absence of electron scattering on impurities (`clean case'), the optical transitions across the SC gap exerted by a uniform light are forbidden. 
The reason is that the hole-like and electron-like states are orthogonal to each other, and thus, they give vanishing matrix elements describing the optical transitions across the SC gap. Soon, the Mattis--Bardeen theory was tested in a number of experiments~\cite{PhysRev.165.588}; it was also generalized to the case of strong electron-phonon interaction~\cite{PhysRev.156.470, PhysRev.156.487} and superconductors with an arbitrary electron mean free path~\cite{ZIMMERMANN199199}.



Nevertheless, the optical transition can still take place in clean superconductors with broken inversion symmetry or in the presence of spin-orbit interaction~\cite{NagaosaOptRe2021}.
Inversion symmetry here can be broken in the presence of a built-in supercurrent~\cite{PhysRevB.106.L220504, PhysRevB.106.214526, PhysRevLett.122.257001}. 
However, the presence of a supercurrent is not a sufficient condition for the optical transitions to occur since the Galilean invariance in parabolic single-band superconductors suppresses the transitions~\cite{Arseev_2006}. 


Two 
ways to break the Galilean symmetry are known: (i) accounting for the non-parabolicity of electronic bands \cite{PhysRevB.95.014506, PhysRevB.106.L220504} and (ii) accounting for the electron-impurity scattering.
Both these scenarios have been realized for the frequencies exceeding the double value of the SC gap.
Recently, it became clear~\cite{PhysRevB.101.134508,SMITH2020168105} that various relaxation time parameters may play an importantant role depending on the ratio between the EM field frequency and SC gap.
It was shown, that at low frequencies, the optical conductivity in the presence of a supercurrent is proportional to the inelastic electron relaxation time, and not to the elastic one. 
At low frequencies and temperatures close to the SC critical temperature $T_c$, the inelastic time is determined by the energy relaxation processes of quasiparticles. 
The energy relaxation time being much larger than the elastic one may result in giant optical conductivity and power absorption in the presence of a supercurrent-carrying state~\cite{PhysRevB.101.134508, SMITH2020168105}.  

%
%
\begin{figure}[t]
\centering \includegraphics[width=0.8\columnwidth]{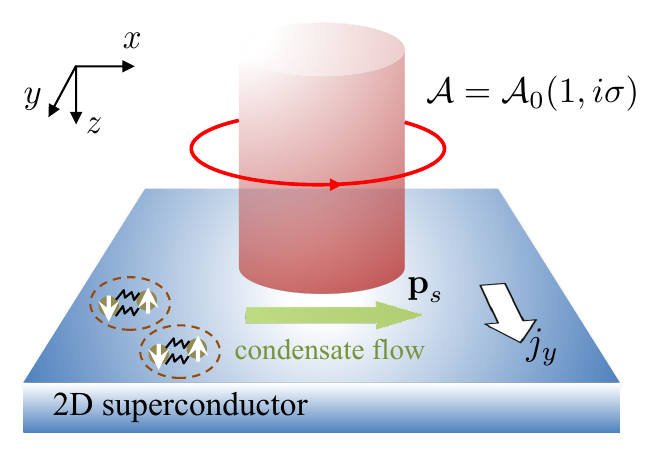} 
\caption{System 
schematic: A two-dimensional (2D) superconductor with a built-in supercurrent  
exposed to an external EM field normally incident to the 2D plane. 
The EM field has circular polarization, characterized by the vector-potential $\mathcal{A}=\mathcal{A}_0(1,i\sigma)$. The built-in supercurrent, which represents a Cooper-pair condensate flow, is aligned along $x$-axis, and it is described by the momentum $\textbf{p}_s$. A photoinduced quasiparticle current $j_y$ emerges in the transverse direction. }
\label{fig1}
\end{figure}

In this Letter, we show that even in a single-band BCS superconductor~\cite{PhysRev.108.1175}, the breaking of both inversion and time-reversal symmetries by means of a built-in supercurrent, and the breaking of Galilean invariance by (weak) electron-impurity scattering results in photoinduced transport of Cooper-pair condensate in the direction transverse to the built-in supercurrent.
Hereby we define the photoinduced anomalous suppercurrent Hall effect. 
At the temperatures $T \ll \Delta$ with $\Delta$ the SC order parameter, the equilibrium density of quasiparticiples above the gap is negligibly small in the absence of external radiation. 
Therefore, we expect that the photoinduced Hall response should be determined by the inelastic quasiparticle relaxation time $\tau_R$ associated with the recombination processes of quasiparticles across the gap. 
It is important to note, that large $\tau_R$ at sufficiently low temperatures provides large values of the supercurrent, opening a way for the experimental verification of its existence.


The idea behind the supercurrent Hall effect can be roughly explained using phenomenological arguments. 
Let us consider a 2D layer with a built-in stationary supercurrent 
generated either by, e.g., a transport current or an external applied magnetic field.
The supercurrent is the consequence of nonzero supermomentum ${\bf p}_s$ of the Cooper pairs, associated with the phase difference of the condensate at the edges of the sample. 
Furthermore, if an isotropic 2D superconductor in supercurrent-carrying regime is normally illuminated by an external EM radiation characterized by the in-plain vector potential ${\bf \mathcal{A}}(t)={\bf \mathcal{A}}\exp{(-i\omega t)}+{\bf \mathcal{A}}^*\exp{(i\omega t)}$, the photoinduced stationary current of quasiparticles excited across the SC gap in the most general form reads as
%
\begin{gather}\label{current}
{\bf j}=a_\omega|{\bf \mathcal{A}}|^2{\bf p}_s+b_\omega[{\bf \mathcal{A}}({\bf \mathcal{A}}^*\cdot{\bf p}_s)+({\bf \mathcal{A}}\cdot{\bf p}_s){\bf \mathcal{A}}^*]\\\nonumber
+ic_\omega[{\bf p}_s\times[{\bf \mathcal{A}}\times{\bf \mathcal{A}}^*]].
\end{gather}
The first term in Eq.~(\ref{current}) gives the longitudinal (aligned along the supercurrent flow) photoexcited current density; the second term contains both the longitudinal and transverse quasiparticle current density responses; the third term gives only the transverse response. 
Anisotropic contributions characterized by the terms proportional to coefficients $b_\omega$ and $c_\omega$ are induced by linearly and circularly polarized radiation, respectively. Note, that 
Eq.~\eqref{current} is valid only for relatively small values of the supercurrent density, $|\textbf{p}_s|v_F\ll\Delta$, where $v_F$ is electron Fermi velocity.

Interested in the Hall transport, we focus on the transverse component of the current~\eqref{current}, $j_y=b_\omega(\mathcal{A}_x\mathcal{A}_y^*+\mathcal{A}^*_x\mathcal{A}_y)p_s+ic_\omega(\mathcal{A}_x\mathcal{A}_y^*-\mathcal{A}^*_x\mathcal{A}_y)p_s$ by choosing the direction of condensate flow along the $x$-axis, ${\bf p}_s=(p_s,0)$, as in Fig.~\ref{fig1}. 
Using the terminology of the two-fluid model, we call $j_y$ the photoexcited current of quasiparticles contributing to the normal component of electron fluid. 
It provides an accumulation of carriers of charge at the transverse boundaries of the sample.
In the case of a non-SC material, such an accumulation results in the emergence of the Hall electric field. 
In the case of a SC material, instead, the electric field cannot penetrate the  SC sample. 
Therefore, the transverse quasiparticle current $j_y$ should be accompanied by an induced \textit{transverse} condensate flow $j_s$ in such a way, that the Hall electric field is compensated, thus $j_s+j_y=0$, and the net transverse electric current vanishes.

Furthermore, even though the net current vanishes, the emergence of $j_s$ produces the condensate phase difference on the transverse boundaries of the sample, 
%
%
$\Delta\phi_H\propto-j_yw$, where $w$ is the width of the 2D SC sample across ${\bf p}_s$ in $y$ direction. 
The Hall-like condensate phase difference $\Delta\phi_H$ directly relates to the coefficients $b_\omega$ and $c_\omega$, which determine the quasiparticle optical response across the SC gap. 
In what follows, let us build a microscopic theory to find $c_\omega$, thus considering circularly polarized EM field.

\begin{figure*}[t]
\centering \includegraphics[width=1.99\columnwidth]{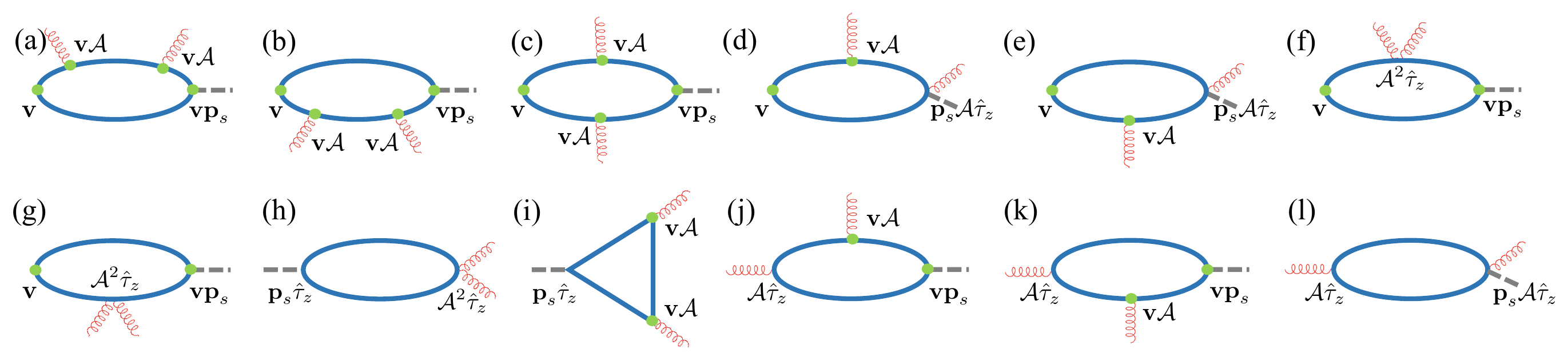} 
\caption{Feynman diagrams describing the photoinduced quasiparticle electric current. 
Blue lines stand for the Green's functions of quasiparticles, red wavy lines indicate the external EM field $\mathcal{A}$, green circles represent the quasiparticle velocity vertex $\textbf{v}$, and dashed lines denote the supercurrent momentum ${\bf p}_s$.}
\label{fig22}
\end{figure*}

\textit{Photoinduced current density.---}
In the absence of relaxation processes, the Hamiltonian of a 2D superconductor with an isotropic $s$-type BCS pairing exposed to an external EM field reads (in $\hbar=k_B=c=1$ units)
\begin{gather}\label{Ham}
\hat{H}=
\left(\begin{array}{cc}
\xi({\bf p}-{\bf p}_s-e\mathcal{A}(t)) & \Delta \\
\Delta  & -\xi({\bf p}+{\bf p}_s+e\mathcal{A}(t))
\end{array}\right).
\end{gather}
Here, $\xi(\textbf{p})\equiv\xi_p=\textbf{p}^2/2m-E_F$ is the electron kinetic energy measured from the Fermi energy $E_F$, and $\Delta$ we assume real-valued. 
The current density operator and the current density obey the standard relations,
\begin{gather}\label{Current}
\hat{{\bf j}}=-\frac{\delta \hat{H}}{\delta\mathcal{A}},~~~~~~~
{\bf j}(t)=-i\,\textmd{Sp}\left\{\hat{{\bf j}}\,\hat{\mathcal{G}}^{<}(t,t)\right\},
\end{gather}
where $\hat{\mathcal{G}}^<(t,t)$ is a lesser component of the Green's function defined by the matrix equation $(i\partial_t-\hat{H})\hat{\mathcal{G}}(t-t')=\delta(t-t')$ in the Nambu and Keldysh representation. 
Expanding Eq.~\eqref{Current} up to the first-order with respect to ${\bf p}_s$ and  to the second order with respect to $\mathcal{A}(t)$ yields a set of Feynman diagrams for the stationary current shown in Fig.~\ref{fig22}.


%
%
%
%
%
%

\textit{Clean case.---} 
%
In the absence of impurities, a single-band superconductor with parabolic electron dispersion possesses the Galilean invariance with or without a built-in supercurrent. 
Consequently, optical absorption vanishes in both cases.
Meanwhile, the second-order stationary response is a consequence of photoabsorption across the gap. 
Thus, it vanishes in clean case both in the absence and presence of the supercurrent. 
The inspection of all the diagrams in Fig.~\ref{fig22} confirms this statement (See Supplemental Material~\cite{[{See Supplemental Material at [URL], which gives the details of the analysis of all the relevant Feynman diagrams}]SMBG}), except one diagram shown in Fig.~\ref{fig22} (l). 
The calculation of this diagram gives a nonzero current density, which seemingly violates the Galilean invariance of the theory. 
To restore the Galilean invariance, one has to account for additional terms reflecting the BCS electron-electron interaction-induced vertex corrections~\cite{PhysRevB.95.014506, SMBG}.

The optical absorption and the photoinduced electric current~\eqref{current} acquire finite values when the Galilean invariance is violated~\cite{PhysRevB.95.014506}. 
We consider the case when it happens due to the presence of electron-impurity scattering in the sample.
Another important ingredient in this problem is the relaxation processes of the photoexcited quasiparticles. 
These processes restrict the infinite accumulation of the photoexcited quasiparticles above the SC gap 
leading to a stationary regime with stationary but nonequilibrium distribution function of photoexcited quasiparticles.


\textit{Impure case.---}
To find the photocurrent in the presence of impurities, we should account for electron scattering and relaxation processes associated with the transition of photoexcited quasiparticles to the SC condensate in the Green's function in Eq.~(\ref{Current}).
The quasiparticle Green's function in the Born approximation averaged over the disorder but in the absence of the EM field and supercurrent reads 
\begin{eqnarray}\label{bareGF}
&&\hat{g}^R_\epsilon(\textbf{p})=\frac{1}{\eta_\epsilon\epsilon-\xi_p\hat{\tau}_z-\Delta\eta_\epsilon\hat{\tau}_x},\\\nonumber
&&\eta_\epsilon=1+\frac{\Theta[\Delta-|\epsilon|]}{2\tau_i\sqrt{\Delta^2-\epsilon^2}}+i\frac{\textrm{sign}(\epsilon)\Theta[|\epsilon|-\Delta]}{2\tau_i\sqrt{\epsilon^2-\Delta^2}}.
\end{eqnarray}
Here, index $R$ stands for `retarded' (advanced Green's function can be found as $\hat{g}^A_{\epsilon}(\textbf{p})=(\hat{g}^R_{\epsilon}(\textbf{p}))^{\ast}$), $\Theta(x)$ is a Heaviside theta-function, and $\tau_i$ is impurity scattering time. 
 
The Green's function~\eqref{bareGF} can be written as a sum of two projections to the electron- and hole-like states as follows,
\begin{gather}\label{GF}
\hat{g}^R_\epsilon=\frac{\hat{A}_p}{\epsilon-\epsilon_p+\frac{i}{2\tau_p}}+
\frac{\hat{B}_p}{\epsilon+\epsilon_p+\frac{i}{2\tau_p}},\\\nonumber
\hat{A}_p=\left(\begin{array}{cc}
u^2 & uv \\
uv  & v^2
 \end{array}\right)+i\gamma \left(\begin{array}{cc}
\frac{1}{2} & uv \\
uv  & \frac{1}{2}
 \end{array}\right)\equiv \hat{A}_0+i\gamma \hat{\Gamma}_A,\\\nonumber
\hat{B}_p=\left(\begin{array}{cc}
v^2 & -uv \\
-uv  & u^2
 \end{array}\right)-i\gamma \left(\begin{array}{cc}
\frac{1}{2} & -uv \\
-uv  & \frac{1}{2}
 \end{array}\right)\equiv \hat{B}_0-i\gamma \hat{\Gamma}_B,\\\nonumber
 u^2=\frac{1}{2}\left(1+\frac{\xi_p}{\epsilon_p}\right),\,\, v^2=\frac{1}{2}\left(1-\frac{\xi_p}{\epsilon_p}\right),
\end{gather}
where $\epsilon_p=\sqrt{\xi_p^2+\Delta^2}$ is a quasiparticles dispersion, $\gamma^{-1}=2\tau_i|\xi_p|$ is impirity-related factor renormalizing the Bogoliubov coefficients $u$ and $v$.
All relaxation processes are characterized by the parameter
\begin{gather}\label{totaltime}
\frac{1}{\tau_p}=\frac{1}{\tau_i}\frac{|\xi_p|}{\epsilon_p}+\frac{1}{\tau_R},
\end{gather}
where the first term in r.h.s. of Eq.~\eqref{totaltime} describes the quasiparticle relaxation due to the scattering off impurities, while the second term accounts for the recombination back to the condensate, characterized by the parameter $\tau_R$. 
In what follows, we assume that under external EM radiation, photoexcited quasiparticles are generated at the gap edge, $2\Delta\gg\omega-2\Delta>0$ and possess the momentum $p\approx p_F$ ($|\xi_p|\rightarrow 0$). 
In this case, the dominant inelastic process is the recombination across the SC gap is characterized by $\tau_R$ since $\tau_p=\tau_i\tau_R\epsilon_p/(\tau_R|\xi_p|+\tau_i\epsilon_p)\approx\tau_R$ at $|\xi_p|\rightarrow 0$. 

At temperatures $T\ll\Delta$ and at the bottom of the quasiparticle branch, the recombination time may acquire large values since~\cite{PhysRevB.39.1602}
\begin{gather}\label{rectime}
\frac{1}{\tau_R}\propto\frac{T}{E_F\tau_i}e^{-2\Delta/T}.
\end{gather}
Indeed, any recombination process represents a transition across the gap after forming the Cooper pair back to the SC condensate. 
The intensity of this process is proportional to the thermal density of quasipaticles above the gap. This density is small at low temperatures, which is reflected by the exponential factor in Eq.~\eqref{rectime}. 

The influence of two types of relaxation processes on the Green's function~(\ref{GF}) is twofold. 
On one hand, both of them shift the Green's function pole in the complex plane (see the imaginary part in the denominator of Eq.~(\ref{GF})). 
On the other hand, the impurity scattering processes give main contribution to renormalization of the Bogoliubov coefficients $u$ and $v$ since the coefficient $\gamma$ is large at $p\approx p_F$ ($\xi_p\rightarrow0$). 
Therefore we replace the matrices  $\hat{A}_0$ and $\hat{B}_0$ by $\hat{A}_p$ and $\hat{B}_p$ in the numerator of Eq.~(\ref{GF}).
It should be noted, that although  $\tau_R\gg\tau_i$, taking into account $\tau_R$ in Eq.~\eqref{totaltime} is still crucial. 
As we will see below, the photoexcited electric current is determined by a large but finite value of $\tau_R$, while the photocurrent diverges in the limit $\tau_R\rightarrow \infty$.


To analyse the current in a superconductor with impurities, we again address the Feynman diagrams in Fig.~\ref{fig22}. 
The diagrams (f)-(i) give only  longitudinal contribution, while the diagrams (a)-(c) result in both longitudinal and transverse response if exposed to linearly polarized EM field. 
The remaining diagrams (d),(e),(j)-(l) in Fig.~\ref{fig22} describe the longitudinal and transverse response in both the cases of linear and circular light polarization. For circularly-polarized light, diagrams (d) and (e) reveal the major contribution in the vicinity of the resonance $\omega\approx2\Delta$
~\cite{[{
It should be noted, that the diagram (l) in Fig.~\ref{fig22} diverges without the account of the impurity-corrected vertices. Accounting for the impurity-ladder corrections results in the regularization of its contribution. Nevertheless, its value is still much smaller than the contributions of diagrams in Fig.~\ref{fig22}(d),(e) in the vicinity of the resonance, see details in~\cite{SMBG}}]SMBG2}.

\begin{figure}
\centering \includegraphics[width=0.95\columnwidth]{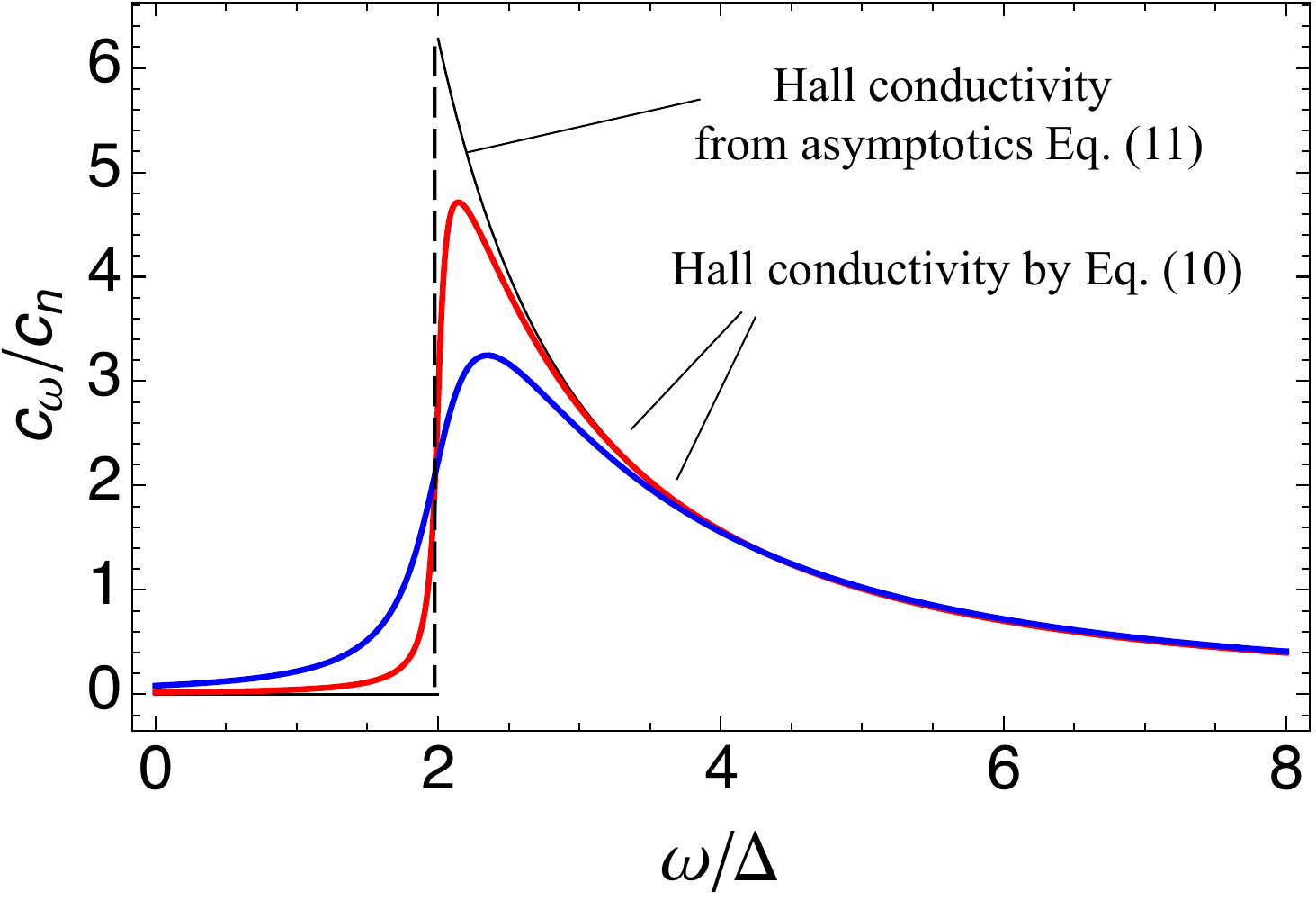}
\caption{Circularly polarized EM field-induced nonlinear transverse conductivity as a function of normalized frequency of EM field (coefficient $c_{\omega}$ from Eq.~(\ref{current}) normalized on $c_n=(e^3/4m\pi)(\tau_R/\tau_i)$ as a function of $\omega/\Delta$ in $\hbar=1$ units).
The calculations were performed using diagram (d) from Fig.~\ref{fig22}.
(Diagram (e) from Fig.~\ref{fig22} gives the same contribution.) 
Blue, red and black curves correspond to
the normalized relaxation times $\tau_R \Delta= 4, 20, \infty$, correspondingly. 
}
\label{plot}
\end{figure}

Let us summarize here the calculation of diagrams (d) and (e)~\cite{SMBG}.
An analytical expression for the current density manifested by panel (d) [diagram (e) gives the same contribution] reads
\begin{gather}
\nonumber
{\bf j}=\frac{ie^3}{m}\sum_{\epsilon,\textbf{p}}\textbf{v}(\textbf{v}\mathcal{A}^{\ast})(\textbf{p}_s\mathcal{A})
\tilde f_{\epsilon,\omega}^{-}
\textmd{Sp}\left\{\hat{g}^R_{\epsilon}[\hat{g}^R_{\epsilon-\omega}-\hat{g}^A_{\epsilon-\omega}]\hat{\tau}_z\hat{g}^A_{\epsilon}\right\}\\
\nonumber
+\frac{ie^3}{m}\sum_{\epsilon,\textbf{p}}\textbf{v}(\textbf{v}\mathcal{A})(\textbf{p}_s\mathcal{A}^{\ast})
\tilde f_{\epsilon,\omega}^{+}
\textmd{Sp}\left\{\hat{g}^R_{\epsilon}[\hat{g}^R_{\epsilon+\omega}-\hat{g}^A_{\epsilon+\omega}]\hat{\tau}_z\hat{g}^A_{\epsilon}\right\},
\end{gather}
where $\tilde f^{\pm}_{\epsilon,\omega}=(f_{\epsilon}-f_{\epsilon\pm\omega})$ with $f_{\epsilon}$ the Fermi-Dirac distribution function, and we omit the momentum dependence of the Green's functions for brevity.
Performing the integration over $\epsilon$ in the limit of zero temperature gives
\begin{eqnarray}\label{Fishcurr12}
{\bf j}=-\frac{2e^3}{m}\sum_{\textbf{p}}\textbf{v}(\textbf{v}\mathcal{A}^{\ast})(\textbf{p}_s\mathcal{A})\frac{\textrm{Sp}[\hat{A}_p\hat{B}_0\hat{\tau}_z\hat{A}^{\ast}_p]}{(2\epsilon_p-\omega)^2+(1/\tau_p)^2}\nonumber\\
+\frac{2e^3}{m}\sum_{\textbf{p}}\textbf{v}(\textbf{v}\mathcal{A})(\textbf{p}_s\mathcal{A}^{\ast})\frac{\textrm{Sp}[\hat{B}_p\hat{A}_0\hat{\tau}_z\hat{B}^{\ast}_p]}{(2\epsilon_p-\omega)^2+(1/\tau_p)^2}.
\end{eqnarray}
Deriving above expression, we used the \textit{resonant} approximation, considering $\omega\approx2\epsilon_p\approx2\Delta$ and keeping only resonant terms in Eq.~\eqref{Fishcurr12} [see the denominators $(2\epsilon_p-\omega)^2+(1/\tau_p)^{2}$ with $\omega>0$]. 
In the framework of this approximation and at zero temperature, $f_{\epsilon=\epsilon_p}\rightarrow 0$ and $f_{\epsilon-\omega\approx-\epsilon_p}\rightarrow 1$. 
Using $\mathcal{A}=\mathcal{A}_0(1, i\sigma)$, where $\sigma=\pm 1$ indicates left/right polarization of EM field, and taking the traces in Eq.~\eqref{Fishcurr12} yields the transverse current density,
\begin{gather}\label{Fishcurr13}
j_y=\frac{8e^3}{m}\sigma p_{s }\mathcal{A}_0^2 \sum_{\textbf{p}}\gamma v_y^2\frac{u^2v^2(u^2-v^2)}{(2\epsilon_p-\omega)^2+(1/\tau_p^2)}.
\end{gather}
The integration over momentum ${\bf p}$ here can be performed in general form~\cite{SMBG}. 
However, the formula can be additionally simplified using the substitution $[(2\epsilon_p-\omega)^2+1/\tau_p^2]^{-1}\rightarrow\pi\tau_p\delta(\omega-2\epsilon_p)$, where $\tau_p\approx\tau_R$ at $p\approx p_F \,(|\xi_p|\rightarrow0)$. 
Restoring the dimensionality, we find the transverse photocurrent, 
\begin{gather}
\label{EqCurrentSimplified}
j_y=\sigma\frac{e^3}{2m\hbar^2}p_{s }\mathcal{A}_0^2 \frac{\Delta^2}{\hbar^2\omega^2}\frac{\tau_R}{\tau_i}\Theta(\hbar\omega-2\Delta),
\end{gather}
which is determined by large parameter $\tau_R/\tau_i$. 
Formula~(\ref{EqCurrentSimplified}) is illustrated in Fig.~\ref{plot} (see thin black line) together with the transverse photocurrent described by Eq.~(\ref{Fishcurr13}) for various $\tau_R$.


\textit{Discussion.---} 
%
%
The theory is fully gauge-invariant in $xy$ plane.
Developing the formalism, we used the gauge $\varphi=0$ for the EM field with a normal incidence to the 2D plane. 
Thus, we considered the second-order response, which is transverse to the direction of EM wave propagation, and none of the collective modes in the SC sample is excited. 
In the case of oblique incidence of the external EM field, the long-range Coulomb forces appear in the sample. 
As a consequence, the gauge invariance requires to take into account collective excitations of the order parameter~\cite{PhysRevB.106.094505, Arseev_2006}.

Furthermore, we assume that the Hall-like condensate phase difference is a consequence of the relation $j_s+j_y=0$ between the photoexcited quasiparticle current $j_y$ and the induced Hall supercurrent $j_s$. 
At the same time, this relation is based on the statement that there is no electric field inside the SC sample. 
A more precise microscopic study shows, that the electric field penetrates into the SC sample up to the distance $\lambda$, which is of the same order as the SC coherence length~\cite{Artemenko_1979}. 
Therefore, our theory is valid for the sample widths $w\gg\lambda$. 
In the opposite case $w\simeq\lambda$, the photoinduced Hall supercurrent becomes spatially-dependent, $j_s=j_s(y)$, across the built-in supercurrent direction in full analogy with the anomalous Hall transport in $p_x+ip_y$ superconductors~\cite{PhysRevB.92.100506}. 
Such a regime requires a separate theoretical investigation. 

Also we assume that the given built-in supercurrent  does not affect the characteristics of the relaxation processes, which play essential role in the theory.
This assumption is valid since we only consider linear in ${\bf p}_s$ photocurrent Eq.~\eqref{current}. 
In principal, the electron-impurity scattering time and the quasiparticle recombination time being scalars may only acquire corrections proportional to $p^2_s$ and higher orders, which is beyond our consideration.


\textit{Conclusion.---} We developed a theory of a photoresponse in a single-band 2D isotropic superconductor with a built-in supercurrent, 
accounting for a random impurity potential, which destroys the Galilean invariance. 
We predicted a photoinduced second-order transport phenomenon -- the emergence of a transverse photoinduced supercurrent, and demonstrated, that its magnitude is primarily determined by the quasiparticle recombination time.
%
The supercurrent Hall effect opens a way to manipulate the direction of superconducting condensate flow via optical tools without external magnetic fields.  
A resent active study of the SC diode effect shows the importance of this phenomenon both from fundamental physics and from the industrial applications perspectives. 
Since the diode effect is fundamental for quantum logic, our work opens perspectives for modeling mesoscopic superconductors-based logical elements.

\textit{Acknowledgements.---} We were supported by the Institute for Basic Science in Korea (Project No.~IBS-R024-D1), Ministry of Science and Higher
Education of the Russian Federation (Project FSUN-2023-0006), and the Foundation for the Advancement of Theoretical Physics and Mathematics ``BASIS''. V.M.K. is grateful to O.V. Kibis for valuable
discussions.

\bibliography{biblio}
\bibliographystyle{apsrev4-2}


\end{document}